 \documentclass{elsarticle}

\usepackage{graphicx}
\usepackage{soul}
\graphicspath{{../figs/}{figs/}{./}}
\usepackage{amssymb}

\usepackage{lineno}

\usepackage{tabularx}

\usepackage[colorinlistoftodos]{todonotes}
\usepackage[outdir=./]{epstopdf}

\makeatletter
\def\ps@pprintTitle{%
 \let\@oddhead\@empty
 \let\@evenhead\@empty
 \def\@oddfoot{}%
 \let\@evenfoot\@oddfoot}
\makeatother

\begin{document}

\begin{frontmatter}


\title{ Predicting Task and Subject Differences with Functional Connectivity and BOLD Variability }



\author[1]{Garren Gaut}
\author[2,3]{Xiangrui Li}
\author[3]{Brandon Turner}
\author[4]{William A. Cunningham}
\author[2,3]{Zhong-Lin Lu \fnref{c2}}
\author[1]{Mark Steyvers \fnref{c1}}

\fntext[c1]{Corresponding author: Mark Steyvers, Department of Cognitive Science, Social and Behavioral Sciences, University of California Irvine, Irvine CA, USA
email: mark.steyvers@uci.edu}
\fntext[c2]{Corresponding author: Zhong-Lin Lu,
Department of Psychology, 
1835 Neil Avenue,
Columbus, OH 43210,
The Ohio State University, Columbus OH, USA
email: lu.535@osu.edu }

\address[1]{Department of Cognitive Science, University of California Irvine, Irvine CA, USA}
\address[2]{Center for Cognitive and Behavioral Brain Imaging, The Ohio State University, Columbus OH, USA}
\address[3]{Department of Psychology, The Ohio State University, Columbus OH, USA}
\address[4]{Department of Psychology, University of Toronto, Toronto ON Canada}
\date{} 

\begin{abstract}
Previous research has found that functional connectivity (FC) can accurately predict the identity of a subject performing a task and the type of task being performed. We replicate these results using a large dataset collected at the OSU Center for Cognitive and Behavioral Brain Imaging. We also introduce a novel perspective on task and subject identity prediction: BOLD Variability (BV). Conceptually, BV is a region-specific measure based on the variance within each brain region. BV is simple to compute, interpret, and visualize. We show that both FC and BV are predictive of task and subject, even across scanning sessions separated by multiple years.  Subject differences rather than task differences account for the majority of changes in BV and FC. Similar to results in FC, we show that BV is reduced during cognitive tasks relative to rest.

\end{abstract}


\end{frontmatter}







 
\section{ Introduction} 

\begin{figure}[h!]
\centering
\includegraphics[width=\textwidth]{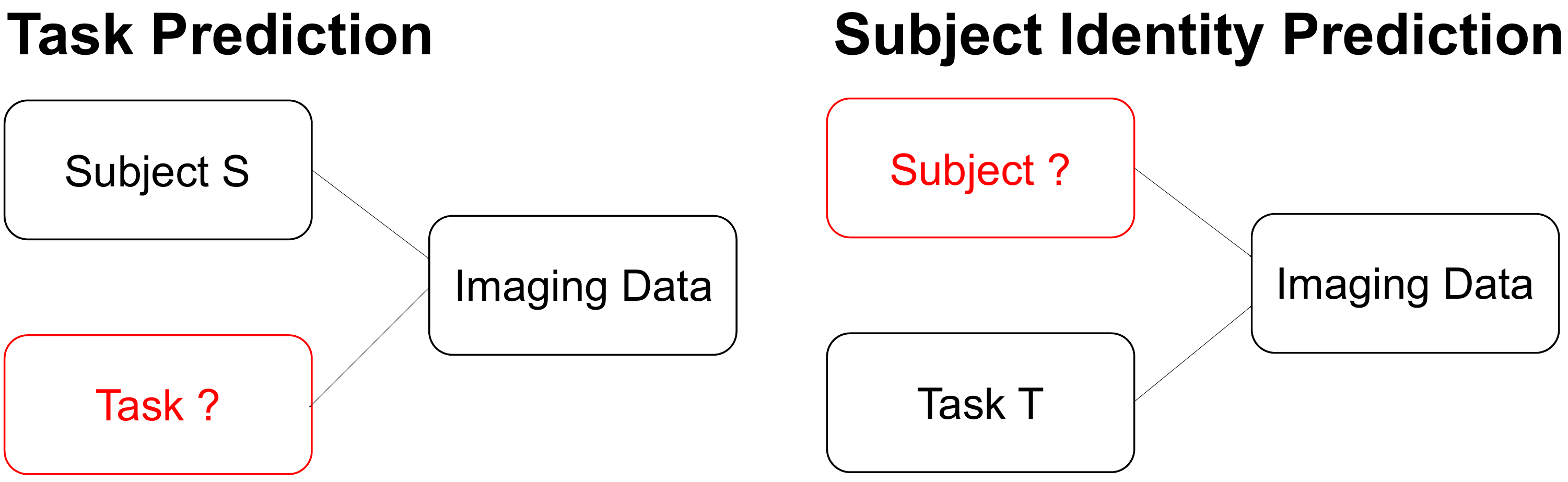}
\caption{ There are two prediction problems addressed in the present article: the task prediction (left) and the subject identity prediction (right). For both settings, an image is generated by a subject performing a task. For task prediction, the goal is to use imaging data to infer which task was performed given the subject's identity. For subject identity prediction, the goal is to use imaging data to infer the unknown subject's identity given the task that was performed.}
\label{prediction_tasks}
\end{figure}

Recent studies have shown that functional connectivity (FC) is highly diagnostic of the task that a subject is performing \citep{Richiardi2011, Shirer2012,GonzalezCastillo2015, Tagliazucchi2012, kucyi2016dynamic, kaufmann2016task}, and the identity of the subject performing a given task \citep{Finn2015,Finn2016psych}. These two prediction problems are illustrated in Figure \ref{prediction_tasks}. For task prediction, a subject is scanned performing an unknown task and the goal is to use imaging data from that scan to predict what task that subject was performing. For subject identity prediction, an unknown subject is scanned performing a known task and the goal is to use imaging data from that scan to predict the subject's identity. The aims of the current work are to replicate results showing that FC is predictive of task and subject identity and demonstrate that a different metric for the variability in BOLD signal, based on BOLD variability (BV), is also highly diagnostic of task and subject identity. 
The present article has three goals: first, to replicate results by showing that FC is predictive of both task and subject identity, second, to demonstrate that BV is also highly diagnostic (albeit less diagnostic than FC) of both task and subject identity, and third, to demonstrate that both FC and BV are robust across preprocessing variations and can be used for subject identity prediction over a timespan of several years.  

\subsection{ Defining FC and BV }

\begin{figure}[h!]
\centering
\includegraphics[width=\textwidth]{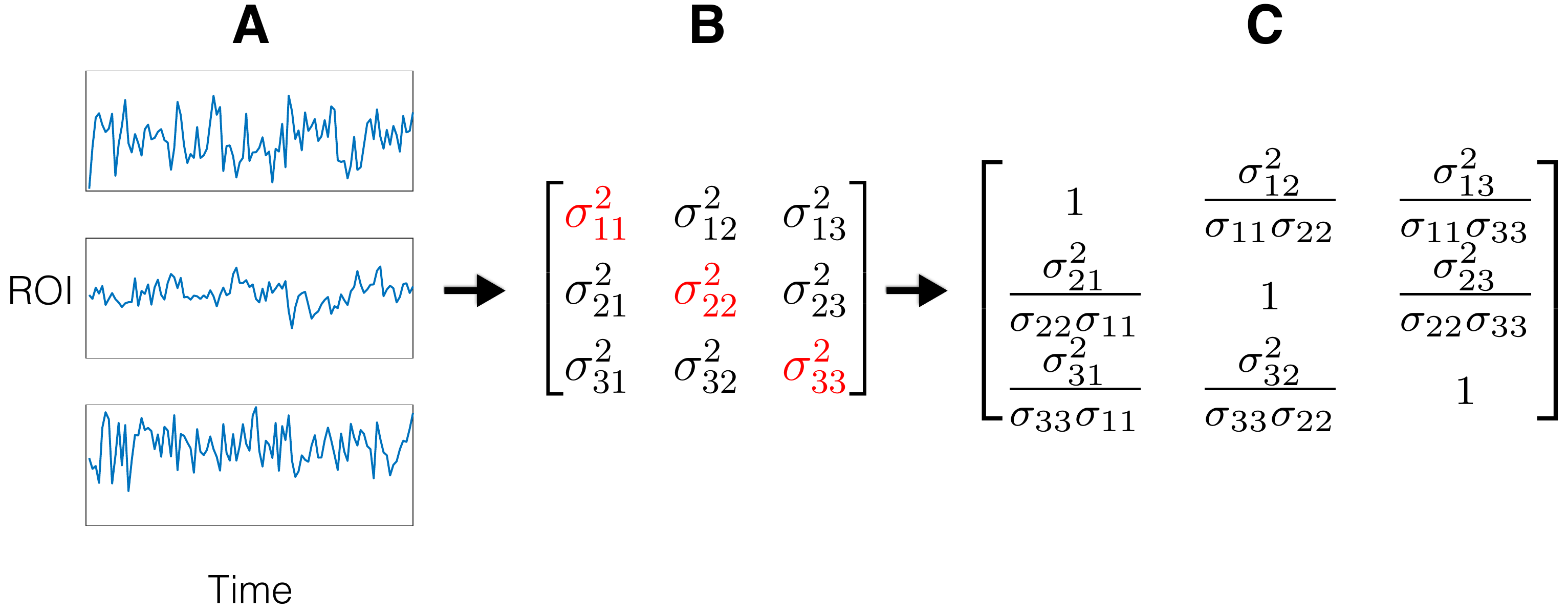}
\caption{ Relations between the calculation of FC and BV. Time series for three ROIs ({\bf A}) are used to compute the covariance matrix ({\bf B}) where  $\sigma_{ij}^2$ represents the covariation between ROIs $i$ and $j$ and the red diagonal entries represent BV. The covariance ({\bf B}) can be used to compute the Pearson correlation matrix ({\bf C}), where the $ij$-th entry of the matrix is $ \sigma_{ij}^2/(\sigma_{ii} \sigma_{jj})$. FC can refer to either the covariance matrix, which explicitly includes BV, or the correlation, which indirectly includes information about the variance. FC is traditionally computed as the correlation and the diagonal of ones is discarded. }
\label{approach}
\end{figure}

FC and BV are two metrics that focus on changes of the BOLD signal around the mean. FC is the correlation or covariance in BOLD activation across regions \citep{friston1993functional, biswal1995functional, Friston2011, VanDenHeuvel2010}. A perspective that might be less familiar to brain connectivity researchers, but is mathematically related to FC, is BV: the region-specific variance in BOLD activation \citep{garrett2010blood,garrett2011importance,garrett2012modulation}. Figure \ref{approach} illustrates how FC and BV are connected. For a given set of ROIs, one can compute the variability of the BOLD time series, and one can compute the degree to which each BOLD time series is related to other time series in the set of ROIs. Both of these calculations are contained in the variance-covariance matrix, where the diagonal elements contain variance terms and the off-diagonal elements contain covariance terms (i.e., Figure 2B). However, because covariance terms are often difficult to interpret, FC is typically based on the Pearson correlation which normalizes the covariance terms by dividing by the variances of the corresponding ROIs, as shown in Figure 2C. As a result, the elements of the correlation matrix are blend of variance and covariance elements, which could potentially obscure important details inherent to tasks, subjects, or ROIs. 


\subsection{ FC predicts task and subject identity }

In the task prediction setting, both whole-brain FC (computed across the entire brain) and network-level FC (computed between subsets of brain regions) have been studied. Whole-brain FC has been shown to accurately predict whether subjects are engaged in a task or at rest \citep{Richiardi2011}, to discriminate between subject-driven cognitive states \citep{Shirer2012}, and to robustly track ongoing cognition \citep{GonzalezCastillo2015}. Furthermore, the ability to track states with FC has been associated with measures of behavioral performance \citep{GonzalezCastillo2015}. FC networks have similarly been shown to predict subject-driven cognitive states \citep{Tagliazucchi2012}, to be associated with attention \citep{kucyi2016dynamic}, and to accurately track task-evoked states \citep{kaufmann2016task}. 

Despite strong links between FC and task-evoked states, recent research suggests that the majority of the variance in FC is accounted for by ``who you are and not what you are doing" (\cite{Finn2016psych} pg. 281). Subjects exhibit individual resting state network architectures that are detectable in task-based fMRI \citep{cole2014intrinsic} and can be used to accurately identify subjects within a group \citep{Finn2015,kaufmann2017delayed,vanderwal2017individual,waller2017evaluating,horien2018considering}. Individual resting state FC has also been used to predict changes in the BOLD signal across task conditions. For example, \citet{Tavor2016} used resting state FC and gross brain morphology to accurately predict BOLD modulation across a range of cognitive paradigms, suggesting that individual differences in task-evoked activity are stable trait markers of underlying individual differences in resting state FC. 

There is no consensus about whether subject-specific FC signatures are persistent across time. One long-term study found that FC within a single individual changed over time and was paralleled by ongoing fluctuations in behavior, although many brain networks are largely stable \citep{poldrack2015long}. Other studies suggest that functional signatures are more stable. For example, \citet{Laumann2015} found that areal parcellation of subject FC is stable over the span of a year, and \citet{choe2015reproducibility} found that resting state FC in a single individual, and especially the executive resting state network, was stable over a three-year period.  

\subsection{ BV associations with task and individual differences}

BV presents a different approach to study BOLD fluctuations that is also associated with task and individual differences. A series of neurocognitive aging experiments \citep[see][for reviews]{garrett2013moment, grady2014understanding} showed age-related effects on task BOLD variability that are separate from, and more predictive than, the mean \citep{garrett2010blood}. A follow-up study \citep{garrett2011importance} identified regions that were associated with age, the speed of response, and consistency of behavioral performance. The difference in variability of high performance-associated regions versus low performance-associated regions was greater for younger, high-performing subjects. In a latent variable study, BV was linked to age, response time, and accuracy in a spatial working memory task \citep{doi:10.1093/cercor/bhv029}. BV in neocortex was also associated with task-related disengagement of the default mode network \citep{doi:10.1093/cercor/bhv029}. BV has also been shown to be related to sub-optimal financial risk tasking among older adults \citep{samanez2010variability}. In addition to age-related effects, individual differences in BV have been associated with lower visual discrimination thresholds \citep{wutte2011physiological}. BV has also been found to vary across task conditions (fixation versus during task) \citep{garrett2012modulation}, and to be associated with task-evoked activity \citep{mennes2011linking}. A study of older adults showed that greater BV was associated with better fluid abilities, better memory, and greater white matter integrity in all white matter tracts \citep{burzynska2015white}.

\subsection{ Goals of the Current Study }

In this study, we replicate results showing that FC is predictive of task and subject identity and test whether BV is also diagnostic of task and subject identity. In this study, we compare the ability of FC and BV in predicting task and subject identity. As FC has already been established as highly predictive of both measures, we aim simply to replicate these results here. However, as BV is studied less frequently, comparing BV to FC provides an important assessment of the relative merits of variance, covariance, and Pearson correlation. We show that connectivity leads to superior results, but that BV can also successfully predict task and subject differences. We also show that the predictive models are robust across time -- FC and BV can be used to predict subject identity across time periods on the order of 3 years, which suggests that subject-specific functional signatures are persistent across time. Finally, we test the robustness of predictive performance across different preprocessing methods. We focus on the preprocessing methods based on the recommended HCP denoising options \citep{burgess2016evaluation}, and investigate the effect of varying changes in the preprocessing pipeline, including the effect of ICA denoising, choice of noise regressors, and whether to regress out the experimental design. 

\section{ Materials and Methods }
\subsection{ Data Acquisition }

MRI recording was performed using a standard 12-channel head coil on a Siemens 3T Trio Magnetic Resonance Imaging System with TIM, housed in the Center for Cognitive and Behavioral Brain Imaging at the Ohio State University (OSU). BOLD functional activations for tasks were measured with a T2*-weighted EPI sequence (repetition time = 2000 msec, echo time = 28 msec, flip angle = 72 deg, field of view = 222 x 222 $\textrm{mm}^2$, in-plane resolution = 74 x 74 pixels or 3 x 3 $\textrm{mm}^2$, 38 slices with thickness of 3 mm). The resting state acquisition had higher resolution (repetition time = 2500 msec, echo time = 28 msec, flip angle = 75 deg, in-plane resolution = 2.5 x 2.5 $\textrm{mm}^2$, 44 slices with thickness of 2.5 mm). T1-weighted structural images were acquired for each subject with the three-dimensional MPRAGE sequence (1 x 1 x 1 mm$\textsuperscript{3}$ resolution, inversion time = 950 msec, repetition time = 1950 msec, echo time = 4.44 msec, flip angle = 12 deg, matrix size = 256 x 224, 176 sagittal slices per slab; scan time 7.5 minutes).

Stimuli were presented to subjects on a rear projection screen through a mirror on top of the head coil. Visual stimuli were generated on a Windows computer running Matlab programs based on Psychtoolbox extensions (http://psychtoolbox.org/).  
The subjects were recruited from the Ohio State University and the surrounding community, and gave informed consent. The experimental protocol was approved by the institutional review board at OSU. A total of 250 subjects participated in the study, but only 174 of them (age 18 to 39, mean 21.6; 63 males and 111 females) were included in the data analysis. A subject was excluded if, during any of the tasks, part of the cerebral cortex was out of the field of view due to head motion, or the mean frame-wise displacement of head motion was greater than 0.15 mm.

During the 1.5-hour MRI session, each subject performed eight behavioral tasks designed to target basic cognitive functions: emotional picture viewing
\citep{cunningham2010aspects}, emotional face viewing \citep{decety2010blame}, episodic memory encoding, episodic memory retrieval \citep{maril2003feeling}, Go/No-go \citep{simmonds2008meta}, monetary incentive \citep{knutson2001anticipation}, working memory \citep{xue2004mapping}, and theory of mind stories/questions \citep{dodell2011fmri}. Resting state scans were also recorded for each subject. Each functional scan lasted about 6 minutes, ranging from 4.1 minutes for the episodic memory retrieval task to 8 minutes for the monetary incentive task. The task descriptions are presented in Table \ref{Tasks} of the Appendix. For convenience of description, the resting state is treated as one of the 9 tasks.

Of the 174 subjects, 19 subjects returned and repeated the experiment approximately 2.8 years (SD=0.4) later. We will refer to this group of subjects as the target group as all machine learning evaluations focus on this group. 

\subsection{ Data Processing }

For fMRI preprocessing, we used the parameters proposed in the minimal preprocessing pipelines of the Human Connectome Project \citep{glasser2013minimal} when applicable. Specifically, the functional brain images were realigned to compensate for head motion, spatially smoothed (2-mm FWHM Gaussian kernel), normalized with a global mean, and masked with the final brain mask. The functional images were then co-registered to the T1-weighted images, and normalized to the standard brain and further refined using nonlinear registration in FSL (FMRIB software library, version 5.0.8, www.fmrib.ox.ac.uk/fsl). Due to spatial resolution of our acquisition, images were not projected to surface space, so the minimal spatial smoothing was performed in volume space.

To denoise the functional data, we followed the HCP FIX-denoising procedures \citep{burgess2016evaluation}, including first lenient high-pass temporal filtering (2000 sec cutoff), motion regression, ICA-based denoising, and mean global time series regression. Additional high-pass filtering (200 sec cutoff) was conducted after regression of the confounding time series.  

\begin{table}
\centering
\caption{Preprocessing variations used to generate each of the four datasets. Each numbered dataset has a single change from the pipeline used to generate the baseline dataset. }
\label{preprocessing_variations}
\begin{tabular}{ c c c c }
\\[-1.5ex] \hline 
{\bf Dataset} & {\bf ICA } & { \bf Regress } & {\bf Noise } \\[-1.5ex] & {\bf denoising} & { \bf Out Design} & {\bf  Regressors} \\ \hline

Baseline  & Yes & No & MGT \\ 
Dataset 1 & Yes & No & CSF+WM \\ 
Dataset 2 & No & No & MGT \\ 
Dataset 3 & Yes & Yes & MGT \\ 
\hline

\end{tabular}
\end{table}

For all datasets, images were parcelled into 299 regions of interest (ROIs) using a functional atlas derived by functional clustering of an outside dataset at the University of Western Ontario \citep{Craddock2012}. 
A mask was used to remove edge voxels to prevent the machine learning classifiers from classifying subjects on the basis of edge-cortex misalignment artifacts created during brain co-registration. To create the mask, we removed any voxels that had low mean intensity in any scan. We removed all ROIs with any voxels that were removed (which is the most conservative approach for removing edge affects, e.g., as opposed to removing ROIs based on a threshold of percentage of voxels removed). The procedure results in 269 ROIs for the subsequent analyses.

\subsection{Feature Generation}
We use the term FC to refer generally to any set of features that requires computing the covariance, and BV to refer to any set of features that requires computing only the variance. For the time series from each task and subject, we compute FC using three different approaches that all depend on entries of the covariance matrix: 1) the Pearson correlation (FCP), 2) the off diagonal entries of the covariance matrix (FCC), and 3) the full covariance matrix (FCCV). FCP and FCC exclude direct information about the variance. However, FCP uses the variance as a normalizing term (see Figure \ref{approach}). We compute BV using two different approaches: the variance (BVV) and the standard deviation (BVSD). 

\subsection{Machine Learning Approach}
Our analysis consists of two prediction tasks: task prediction and subject identity prediction. The goal of task prediction is to predict which task a test subject was performing during scanning given features computed from the scan (random performance in this task amounts to 1/9 = 11\% accuracy). The goal of subject identity prediction is to predict which subject generated a test scan given features computed from the scan (random performance in this task amounts to 1/174 = 0.57\% accuracy). Task prediction and subject identity prediction are evaluated in two settings: within-session and between-session. For within-session prediction, all training and test data are taken from session 1. For between-session prediction, training data are taken from session 1 and test data are taken from session 2. For task prediction, we exclude session 1 scans of the target group from training so that the classifier learns from only task-related (i.e., not subject-related) information. Because fewer subjects participated in session 2, we restrict test sets to only the target group (i.e., 19 subjects that were scanned in both sessions 1 and 2), allowing us to directly compare within-session and between-session performance. However, note that for the subject identification task, the models were not informed of this restriction and have to discriminate between all 174 subjects who participated in the experiment. 

To be consistent with previous analyses \citep{Finn2015}, we used multinomial logistic regression (LR) for task prediction, and a nearest neighbor (1-NN) model for subject identity prediction. The models are evaluated differently as specified in the next section. 

\subsubsection{ Multinomial Logistic Regression } 

Regularized multinomial logistic regression models \citep{hoerl1970ridge,Tibshirani, zou2005regularization} learn to discriminate between multiple class labels for a given data point. Feature weights are regularized using a choice of norm (L1, L2, elastic) and a parameter $\lambda$ that controls the strength of regularization. We use an L2 penalty optimize $\lambda$ over the set $\{ 1^x, x \in \; \{-10, -9, ..., 3 \} \}$.
Regularization usually results in improved generalization performance and is important in our analysis because it allows us to fit models using FC feature sets where the number of features is larger than the number of data points. We used LIBLINEAR \citep{LIBLINEAR} to fit all logistic regression models. For each prediction task (task and subject identity prediction), we trained independent models.

We used stratified nested cross validation 5 outer folds and 2 inner folds to evaluate and select models.  The cross validation procedure was stratified in order to guarantee that a particular test subject always had some data used for training. We evaluate models using out-of-sample accuracy and choose the model with the highest accuracy.


\subsubsection{Nearest Neighbor Model} 
In contrast to LR models that learn from information across tasks, our 1-NN models were restricted to information from pairs of tasks where one was used for test and the other could be thought of as a training set. In principle, the 1-NN model could be set up analogously to the LR model, but we replicated analyses used in previous work \citep{Finn2015} that were used to investigate whether functional signatures indicative of subject identity are preserved across pairs of tasks. 

Each 1-NN model took as input a test instance from task A and a set of labeled training instances from all subjects in task B, where each instance was comprised of features computed from a scan from a particular subject in a particular task. The predicted identity was the identity of the subject corresponding to the nearest training instance, where we defined similarity using the Pearson correlation. For between-session prediction, we iterated through all pairs of tasks A and B. For within-session prediction, we excluded pairs consisting of the same tasks (e.g., A-A) because each task was performed only once per session. To give a comparable set-up to LR task prediction, the test instances were always chosen from the target group and training instances were always chosen from session 1.

\section{ Results } 

Data and analyses reported in this article are publicly available on the Open Science Framework (https://osf.io/gcw4x/). First, we examine patterns in BV organized by subjects and tasks. Next, we show the performance of the machine learning classifiers using the baseline dataset (ICA, experimental design included, MGT as noise regressors). We contrast the relative diagnosticity of BV and FC in the three prediction tasks. Finally, we overview the impact of different preprocessing options on prediction results. 

\subsection{ Visualizing BOLD Variability } 
\begin{figure}[h!]
\centering
	\includegraphics[width=\textwidth]{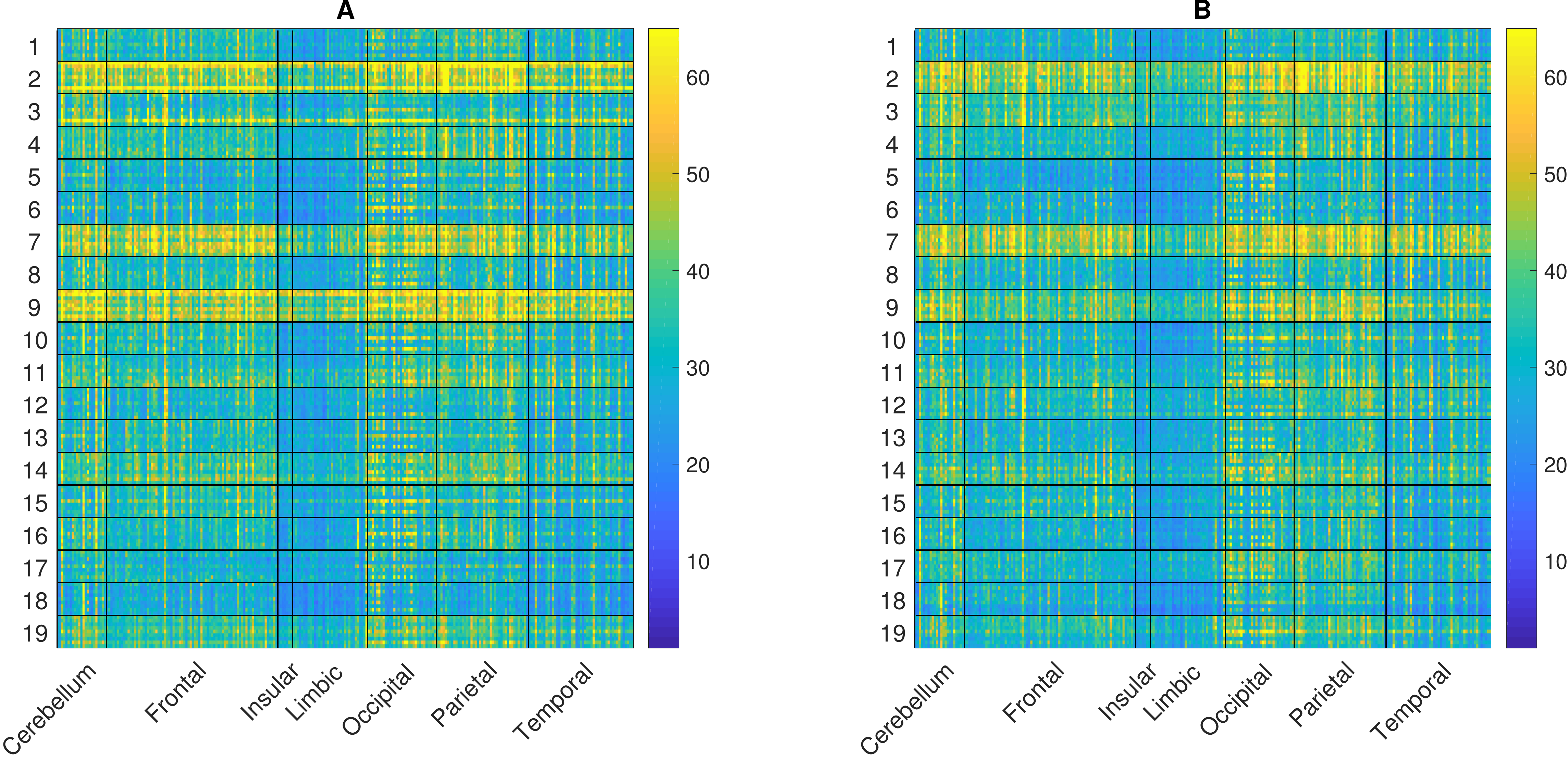}
	\caption{ BV for 19 subjects from session 1 ({\bf A}) and session 2 ({\bf B}). The y-axis organizes scans first by subject and then by task. The x-axis organizes ROIs first by lobe and then ROI. Note that BV is computed by BVSD. } 
\label{19subjectheatmap_subj}
\end{figure}

Figure \ref{19subjectheatmap_subj} shows BVSD for the target group of subjects in session 1 (panel A) and session 2 (panel B). Rows are first grouped by subject and then by task. Columns are first grouped by brain lobe then by ROI. The results show subject-specific patterns in BV that are preserved between sessions. For example, subjects 2, 7, and 9 have relatively high BV in both sessions regardless of task, and subject 18 seems to have relatively low BV in both sessions regardless of task. None of these subjects with outlying BV were outlying in demographic categories (weight, age, height, race). Subjects 2, 7, and 9 have higher mean frame displacement than the other subjects (mean of 0.21 versus 0.09). Subject 5 has relatively low Frontal BV, but average Occipital and Parietal BV. In addition, the results show lobe-specific effects that are also preserved between sessions. For example, Limbic BV is on average lower than Parietal BV. The variance in BV between regions in the Occipital lobe is higher than in other lobes. 

\begin{figure}[h!]
\centering
	\includegraphics[width=\textwidth]{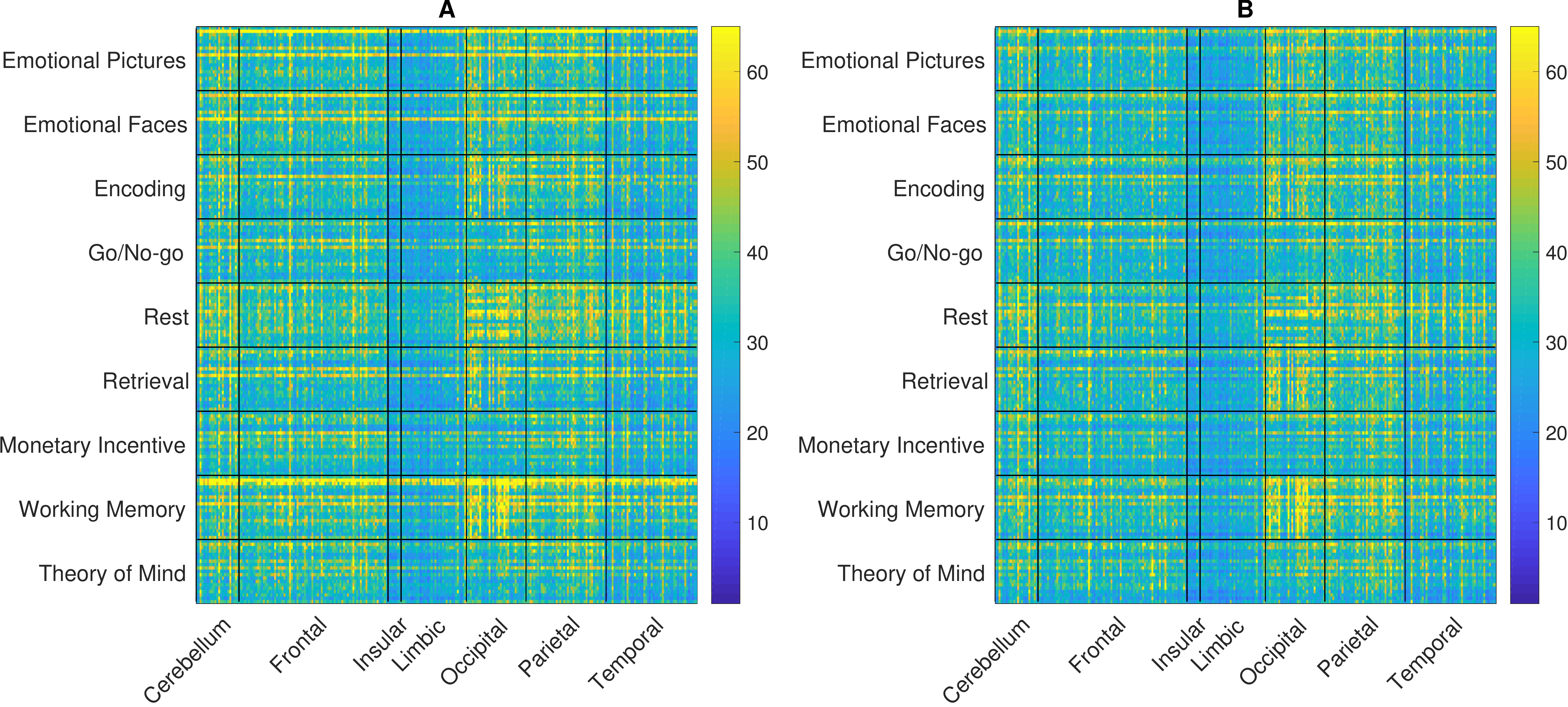}
	\caption{ BV for 19 subjects from session 1 ({\bf A}) and session 2 ({\bf B}). On the y-axis are scans ordered by task. Within each task, scans are ordered by subject. On the x-axis are ROIs ordered by lobe of the brain. } 
\label{19subjectheatmap_task}
\end{figure}

Figure \ref{19subjectheatmap_task} shows BV (computed by BVSD) for the target group subjects in session 1 (panel A) and session 2 (panel B). Rows are first grouped by task and then by subject. Columns are first grouped by brain lobe then by ROI. When ordered by task, BV shows patterns that are preserved across session (e.g., lobe-specific or task-specific effects). For example, Occipital activation is higher for the Theory of Mind task, and Temporal activation is higher during resting state. Aside from these two effects, based on visual inspection of Figure \ref{19subjectheatmap_task}, the BV patterns do not seem to be task specific. However, the machine learning models (discussed in the next section) will demonstrate that the patterns contain diagnostic information that distinguishes the tasks.

 \begin{figure}[h!]
\centering
\includegraphics[width=.66\textwidth]{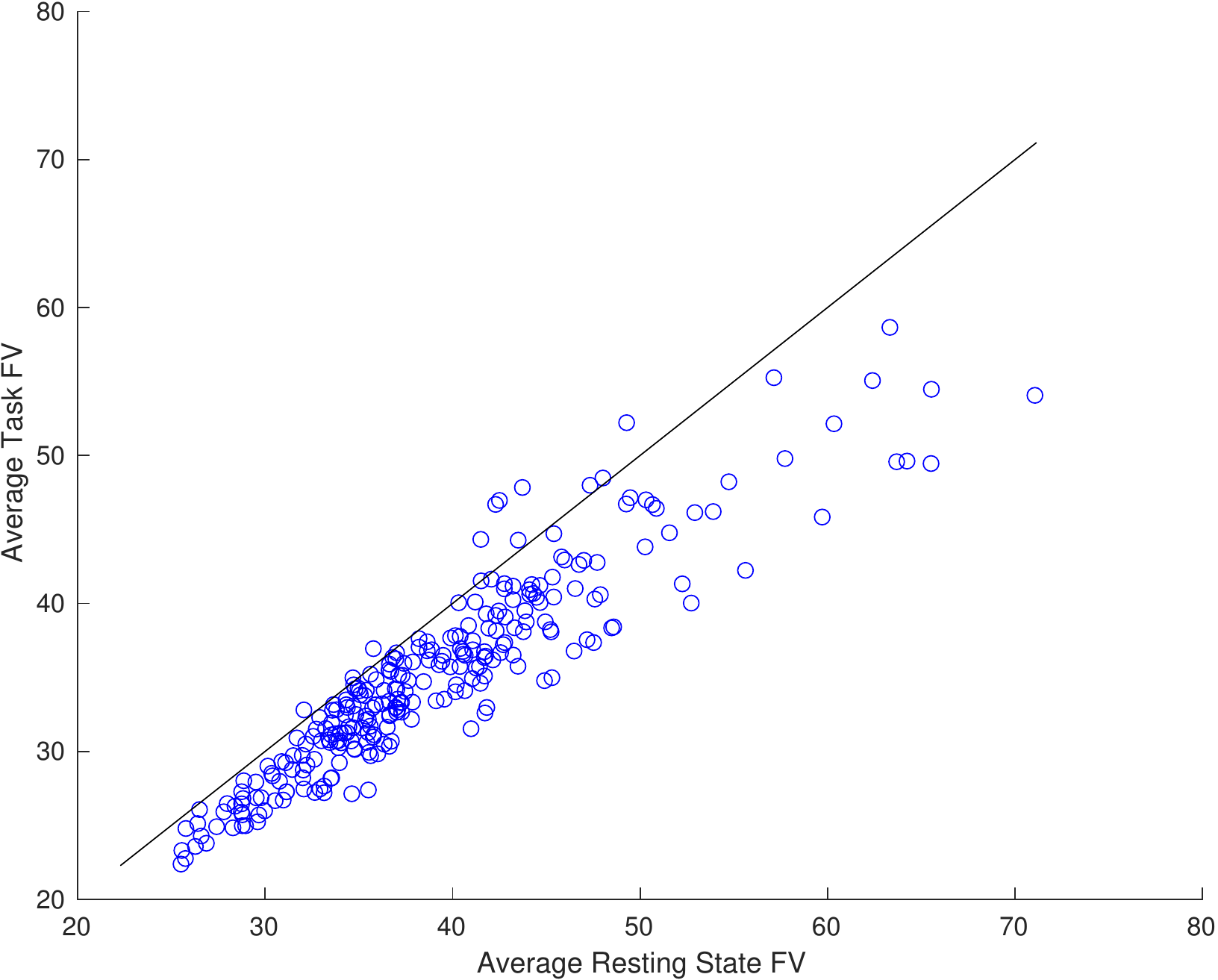}
\caption{BOLD variability in a resting-state task versus non-resting cognitive tasks. Each point represents an individual ROI (averaged over subjects) and the reference line indicates equal BV in resting and non-resting state tasks. } 
\label{RSvTask}
\end{figure}

Finally, we can re-examine known effects of FC through the lens of BV. For example, research has shown a reduction of covariance in the default mode network during task compared to rest \citep{greicius2003functional}. We examine whether this result can be extended to BV. Figure \ref{RSvTask} shows resting state BV versus non-resting state BV in each ROI averaged over all subjects and tasks. Analogous to the effect in FC, for almost all ROIs, resting state BV is higher than task BV.




\subsection{Task Prediction} 
Task out-of-sample LR prediction accuracy is reported in Table \ref{mlresults_LR}. Overall, BV and FC accurately predict task. All models show performance well above chance (1/9=11\%) for all feature sets. However, there is a clear performance benefit when using FC versus BV. In addition, within-session performance is consistently better than between-session performance across all feature sets. The difference between within-session and between-session accuracy is lower on average for BV (7.5\%) versus FC (14\%), suggesting that FC contains more session-specific information than BV.  The particular method of computing FC does not strongly affect predictive performance but there is some suggestive evidence that BVSD leads to more accurate predictions than BVV. 
 \begin{table}
\centering
\caption{Predictive accuracy (percentage correct) of the Logistic Regression model for task classification for different methods of computing functional connectivity (FC) and BOLD variability (BV) and method for assessing generalization (within or between scanning sessions). The 95\% credible interval is reported in parenthesis. } 
\label{mlresults_LR} 
\small
 \begin{tabular}{ l l l l l  } 
{\bf Type} & {\bf Feature}  & {\bf \# Features} & {\bf Within } & {\bf Between }\\ \hline \\[-2ex] 
BV & BVSD & 269  &  79 (72, 84) &  70 (62, 76) \\ 
 BV & BVV & 269  &  66 (59, 73) &  60 (53, 67) \\ 
 FC & FCP & $\frac{269*268}{2}$  &  95 (90, 97) &  83 (77, 88)    \\ 
 FC & FCC & $\frac{269*268}{2}$  &  92 (87, 95) &  82 (75, 87)  \\ 
 FC & FCCV & $\frac{269*269}{2}$  &  95 (91, 98) &  84 (78, 89) \\ 
  \\[-1.5ex] \hline 
 \end{tabular}
\end{table}

\begin{figure}[h!]
\centering
\includegraphics[width=\textwidth]{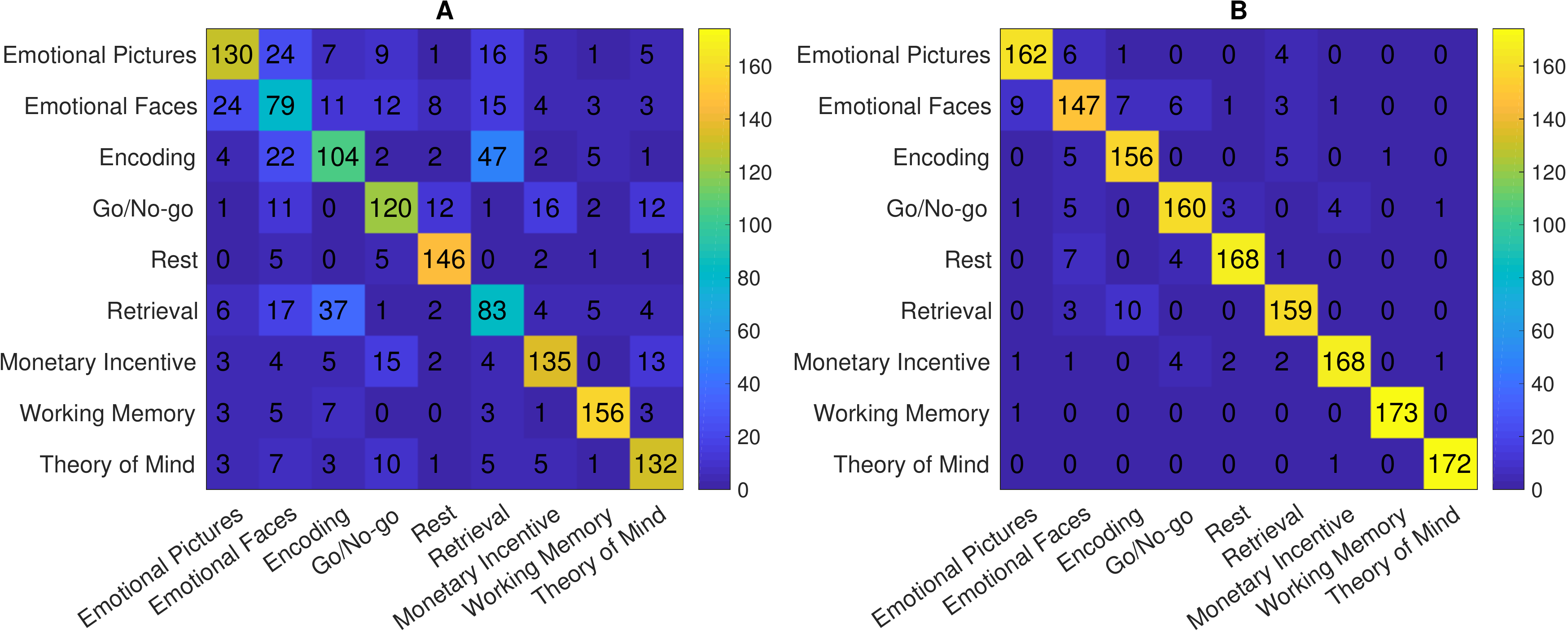}
\caption{ Confusion matrices for task prediction using BV (panel {\bf A}) and FC  (panel {\bf B}). The y-axis corresponds to true task and the x-axis to predicted task. BV and FC were computed using the BVSD and FCP methods respectively}
\label{confusionpng}
\end{figure}

In order to understand the relative performance differences between BV and FC, we compare the confusion matrices in Figure \ref{confusionpng}. Some cognitive tasks  are more difficult to discriminate on the basis of BV. For example, the Emotional Faces and Emotional Pictures tasks (both involving emotional processing despite different visual inputs) and Encoding and Retrieval tasks (both involving episodic memory) are occasionally confused on the basis of BV but less so for FC, suggesting that FC contains unique information that discriminates between these tasks. For a number of tasks (e.g., Resting State or Theory of Mind) discrimination using BV is comparable to FC. The FC model makes more errors than the BV model for only a few cells in the confusion matrix (e.g., Rest-Emotional Faces, and Rest-Retrieval). There is only one cell for which the FC model makes an error where the BV model does not (Rest-Retrieval). Additionally, the structure of errors is similar between BV and FC (i.e., the two models tend to make errors on similar pairs of tasks). The non-diagonal elements of the confusion matrices have a Pearson correlation of 0.75, suggesting that BV and FC make similar types of errors, but that BV makes those errors more often.

\subsection{ Subject Identity Prediction } 
Out-of-sample subject prediction accuracy and 95\% credible intervals for the 1-NN models are reported in Table \ref{mlresults_NN}. Subject identity performance is high regardless of features used (chance performance is 1/174 = 0.57\%). There is overlap in credible intervals for overall accuracy for all feature types  except for BV computed as BVV, which performs significantly worse than the other methods examined. There is no significant overall performance advantage for any of the other methods used. 

There are between-session performance differences based on whether the training and test images were recorded from the same task, from different tasks, or from rest. For all methods used to compute BV and FC, same task to same task accuracy is significantly higher than different task to different task accuracy (95\% credible intervals do not overlap). For all methods but one, FC computed as the Pearson correlation (FCP), accuracy is significantly higher for same task to same task prediction compared to rest to rest prediction. Accuracy for different task to different task prediction is not significantly different than rest to rest accuracy. We suspect that given more rest  to rest observations this difference would become significant; the relatively low number of rest to rest outcomes (one per subject) as compared to different task to different task outcomes (72 per subject) or same task to same task outcomes (8 per subject) leads to larger rest to rest credible intervals. 

\begin{table}[h!]
\centering
\caption{  Subject classification predictive accuracy (percentage correct) and 95\% credible intervals for the Nearest Neighbor models  using different methods of computing functional connectivity (FC) and BOLD variability (BV). For each model accuracy is tested within-session and between-session. For between-session accuracy we report whether the training and test image were selected from the same task, different task, or from rest. } 
\label{mlresults_NN} 
\footnotesize
  
 
 \centering
\begin{tabular}{ l l l l l l l } 
 \\[-1ex] \hline \\[-2ex] {\bf Method}  & {\bf \# Feat} & {\bf Within } & \multicolumn{4}{c}{\bf Between } \\ \cline{4-7}
 & & & {\bf All} & {\bf Same Task} & {\bf Diff Task} & {\bf Rest} \\
BVSD & 269  &  83 (81, 85) &  70 (67, 72) &  93 (88, 96) &  67 (65, 70) &  58 (36, 77) \\ 
BVV & 269  &  73 (70, 75) &  54 (52, 57) &  83 (76, 88) &  51 (49, 54) &  42 (23, 64) \\ 
FCP & $\frac{269*268}{2}$  &  83 (81, 85) &  63 (61, 66) &  80 (73, 86) &  62 (59, 64) &  53 (32, 73) \\ 
FCC & $\frac{269*268}{2}$  &  81 (79, 83) &  63 (60, 65) &  85 (78, 90) &  60 (58, 63) &  47 (27, 68) \\ 
FCCV & $\frac{269*269}{2}$  &  83 (81, 85) &  67 (64, 69) &  86 (79, 90) &  65 (62, 67) &  58 (36, 77) \\ 
 \\[-1.5ex] \hline 
 \end{tabular}

\end{table}

\subsubsection{ Pairwise Subject Identification Accuracy} 
\begin{figure}[h!]
\includegraphics[width=\textwidth]{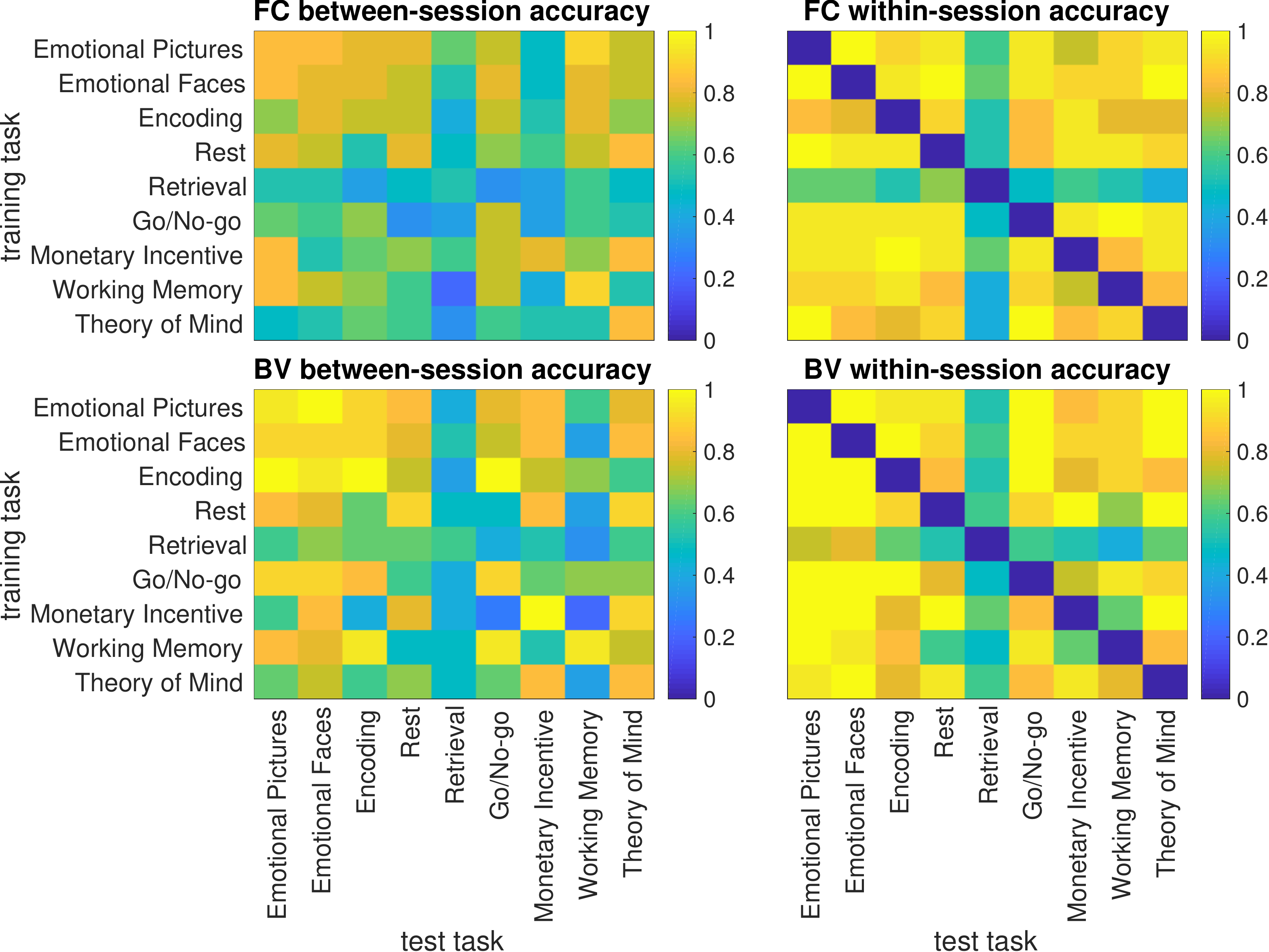}
\caption{ Heatmaps of between and within-session average subject identification accuracy ordered by task. The x-axis shows the task from which test scans were taken and the y-axis shows the task used to predict subject identity.}
\label{task2taskacc}
\end{figure}

The primary purpose of using nearest neighbor models is to build upon past results by Finn et al. 2015 who investigated how subject-specific FC is preserved across tasks. Figure \ref{task2taskacc} shows subject identity prediction accuracy as a function of training and test task. The x-axes show the training task and the y-axes show the test task. Prediction accuracy is averaged over subjects. The top panels and the lower panels correspond to FC and BV, respectively. The left panels and right panels correspond to between-session and within-session, respectively. Within-session accuracy is higher than between-session accuracy for both FC and BV, which is consistent with classifier results in Table \ref{mlresults_NN}. For between-session prediction, performance is best when generalizing between the same tasks. 
    
In the previous framework, Finn et. al. compared nearest neighbor subject identity prediction performance across sessions using FCP. They used three types of train/test pairings: FCP from resting state, FCP from rest and another task, and FCP from two different tasks. They found that rest-to-rest prediction is most accurate (92.9\%) and that accuracy rates ranged from 54.0\% to 87.3\% for other database and target pairs, including rest-to-task and task-to-different-task comparisons. Our analysis includes the additional setting where the training and test sets are from data recorded from the same task. We find that FCP prediction performance is highest for the task-to-same-task setting (80\%), followed by task-to-different-task (62\%), and rest-to-rest (53\%). Overall, our average accuracy is lower (65\% versus 82.1\%), but the performance difference could be attributed to longer time between scanning session (2.8 years versus 2 days) and incorporating non-discriminative data (data from 155 non-target group subjects) into our framework.


\section{Discussion} 

Using two supervised machine learning approaches, we have shown that both BV and FC significantly predict task and subject differences above chance levels. We have also shown that the predictive models are robust across time periods on the order of 3 years, suggesting that subject- and task-specific FC and BV signatures are persistent across time. While FC leads to better predictive performance, there are two features of the BV approach that are worth highlighting. First, BV has a straightforward interpretation: BV is computed from each single brain region, and task- or subject-related BV changes can be attributed to specific brain regions. Second, BV is low dimensional, which leads to simpler computation, easier data visualization, and application of advanced modeling techniques (e.g., fully Bayesian Inference) that can be challenging for more complex brain imaging measures such as FC. 


It is possible that our machine learning models could not adequately capture the valuable information present in FC given the relatively small amount of training data. While a larger training set and more advanced models would improve the performance of both FC and BV, we expect that these changes would better leverage the high dimensionality of FC and lead to better performance gains than BV. Chen and Hu 2018 used a recurrent neural network to improve subject identification performance using short scans and we think that similar methods could be used to improve our modeling results. However, because the goals of this work were to compare the information contained in BV and FC, and not prediction in itself, we focused on simple machine learning models.


Our nearest neighbor prediction framework mirrors past analyses \citep{Finn2015,waller2017evaluating,horien2018considering} that investigated the persistence of subject-specific FC signatures across pairs of tasks, but extends the framework by examining task-to-same-task prediction. Our analysis found that, for both FV and BV, task-to-same-task prediction performed best, and that rest-to-rest prediction performed worst. These results suggest that task engagement modulates the uniqueness of subject-specific BOLD responses in a way that increases subject discriminability. The lack of this modulatory task effect in Finn et. al. could be explained by their different parcellation scheme. Whereas the parcellation method used in Finn et. al. focused on preserving individual connectivity, our graph-based method smoothed functional information more across individuals. As in Waller et al. 2017, our dataset lead to reduced accuracy compared to subject identification for the HCP \citep{Finn2015}. Lower relative performance could be accounted for by reduced spatiotemporal resolution \citep{waller2017evaluating}, since scans from our dataset are about the same length as HCP scan lengths (~6 minutes), and we used motion correction, censored outlying time points, and performed analysis on only low motion subjects (< 0.15 mm average frame displacement), we don’t believe scan length nor head motion are the primary causes of reduced performance.

When we directly contrasted performance in subject identity prediction and task prediction, we found that out-of-sample subject identity prediction was more accurate than task prediction, even though a priori the subject identity task is a more challenging task (i.e., identifying 1/174 versus 1/9). This provides further evidence that ``the majority of the variance in [functional signature] is accounted for by who you are and not what you are doing" \citep{Finn2016psych}. A recent study of task and subject FC expanded upon this idea by showing not only was FC individuality a predominant factor in group-level FC variability, but that task sensitivity could be improved by removing subject connectivity \citep{xie2017whole}. 

There is debate in the field about whether subject-specific functional signatures are persistent across time. One long-term study found that FC within a single individual changed over time and is paralleled by ongoing fluctuations in behavior, although many brain networks are largely stable \citep{poldrack2015long}. Other studies found that parcellation of subject FC is stable over the span of a year \citep{Laumann2015}, and that resting state FC in a single individual, and especially the executive resting state network, was stable over a three year period \citep{choe2015reproducibility}. Our results show that FC and BV can be used to predict subject identity across time periods on the order of 3 years, and suggest that subject-specific functional signatures are persistent across time. 

One potential concern is that past research showed that vascular effects are present in motor tasks and to a much lesser extent, cognitive tasks \citep{kannurpatti2010neural}. This research suggests a potential confound for our results:  vascular effects, rather than neural effects, lead to high predictive performance. We believe that this is not the case for the following two reasons. First, there were only moderate motor components to the cognitive tasks used in our experiment; the only motor components involved reporting answers using button presses. We can expect the vascular effects due to motor control to be less for these tasks compared to the finger tapping task used in the Kannurpatti et al. study. Second, the similarity between motor components of each task (i.e., infrequent button pressing), suggests that even if large vascular effects were present, these effects alone would not be sufficient to discriminate between 9 separate tasks. 

Another potential concern is that structural information contributes to the predictive performance in a way that is separable from functional information. It is possible that differences in gross brain morphology create artifacts in functional signatures during the registration process \citep{jenkinson2001global} that affect both FC and BV measures. For the goal of predicting what cognitive task a subject is engaged in, only functional information can be used to distinguish between cognitive tasks. Therefore, the ability of the model to identify tasks demonstrates that both BV and FC contain diagnostic functional information, and that these functional signatures persist over time. For the subject identity prediction task however, care has to be taken in interpreting the results. The identification of a person based on structural information is not an impressive outcome compared to the identification of a person based on functional connectivity or functional variability. For this reason, we did not use LR models for subject identification because they could easily overfit to a structural confound. The 1-NN classifier does not make use of any free parameters that can be tuned to particular ROIs, and therefore the identification occurs on the basis of overall similarity between functional signatures and not any particular ROI. During preprocessing, we carefully ensured that high subject identification performance was not due to structural confounds: we removed edge voxels (i.e., those most likely to be misaligned) from our analysis, used non-linear registration, and performed separate registration for each scanning session. Furthermore, brain parcellation was performed using a dataset from a separate population, which reduces the probability that voxels were grouped into regions that a priori differentiate subject identity (i.e., that ROIs reflect subject-specific rather than task-specific functional differences). Therefore, even if structural information affects particular ROIs, it is unlikely that the classification results in the subject identification task are driven entirely by structural information. However, future research will investigate how structural information might contribute to classification performance. 

\section{ Conclusions } 

Our results replicate the general findings that FC is predictive of both task and subject. We have shown that simple statistics like BV also can capture information that is diagnostic of both task and subject. The predictive performance of both FC and BV are robust across time and across common variations in preprocessing. 


%

\section{Funding}
       \vskip1ex
       This work was supported by a National Sciences Foundation Integrative Strategies for Understanding Neural and Cognitive Systems Collaborative Research Grant (1533500 and 1533661).


\section{ Supplementary Material }

\subsection{ Ensuring Robustness of FC and BV Results}
To ensure that FC and BV results are robust, signal due to head motion and physiological processes must be removed, and the data must be denoised. 
Denoising can be done by censoring high-motion time points \citep{siegel2014statistical}, independent component analysis \citep[ICA;][]{salimi2014automatic, griffanti2014ica}, regressing out estimated motion parameters \citep{friston1996movement}, regressing out physiological measurements \citep{glover2000image}, or regressing out nuisance regressors derived from brain regions associated with physiological noise \citep{behzadi2007component}. A recent study compared motion denoising strategies, censoring, ICA, motion regression, and regression of the mean grayordinate time series (MGTR), and found that a combination of ICA and MGTR were the best methods for reducing noise due to motion \citep{burgess2016evaluation}. However, the effect of different pre-processsing pipelines on prediction studies has not been tested. 

\subsubsection{Preprocessing Variations}

To explore the effects of preprocessing options on the results, we created four different preprocessing variations from the HCP pipeline. The pipeline differed on 
{\it i}) whether or not to perform Independent Component Analysis (ICA) denoising, {\it ii}) whether or not to regress out the experimental design, and {\it iii}) to regress out cerebrospinal fluid and white matter (CSF and WM) or to perform mean global time series (MGT) signal regression \citep{burgess2016evaluation}. 

We used FIX to perform ICA-based denoising \citep{salimi2014automatic, griffanti2014ica}. We selected training noise components in a conservative way; we considered components with very high or very low frequencies and on the edge of the brain to be noise. 

The effect of regressing out the experimental design matrix corresponds to a prediction scenario where only the variance around the mean trend (e.g., the residual variance) is used. While removing the mean BOLD trend may lower predictive performance, we find that this scenario is important for testing the predictive performance when BV and FC are characterized as orthogonal to the mean trend. When the experimental design is not regressed out, we test the predictive performance using features that include experimentally related changes in the mean trend. In this scenario, we don't assume a model of the BOLD time series mean response, i.e., BV and FC are influenced by changes in mean trend.  

Finally, we also tested whether the choice of noise regressors impacts predictive performance. We compare the effects of physiological noise regressors (CSF and WM) to the mean global time series (MGT) regressor. This is similar to mean grayordinate time series regression, which has been shown to greatly reduce artifacts related to whole-brain motion \citep{burgess2016evaluation}. 

Each variation of the pre-processing pipeline generated a new dataset. To keep the number of datasets manageable, we began with a baseline dataset and created four separate datasets where we changed only a single preprocessing step at a time (see Table \ref{preprocessing_variations}). The baseline dataset follows the recommended HCP processing pipeline \citep{burgess2016evaluation} and includes ICA denoising, regresses out the mean global time series, and does not regress out experimental design. 

\subsubsection{Robustness of task classification results across preprocessing options}

Predictive performance for task classification was not significantly different between any of the preprocessing options we examined. Performance differences were extremely small when {\it i}) using ICA vs. not and {\it ii}) regressing out CSF and WM versus regressing out the MGT. Results were better when using ICA vs. No ICA. The maximum pairwise (by feature type) differences in accuracy between ICA and No ICA were $\approx$ 2.33\% and 2.92\%, for within-session and between-session prediction, respectively. These differences were about the same for BV and FC. 

For regressing out MGT vs. CSF + WM, differences in accuracy were small and did not consistently favor one option over the other. The maximum pairwise (by feature type) differences in accuracy between the dataset comparing mean global regression vs. CSF+WM regression were $\approx$ 1.6\% and 6.43\%, for within-session and between-session prediction, respectively. Differences were larger for BV ($\approx$ 4.67\% and 6.43\%) compared to FC ( $\approx$ 1.16\%). 

\begin{figure}[h!]
\includegraphics[width=\textwidth]{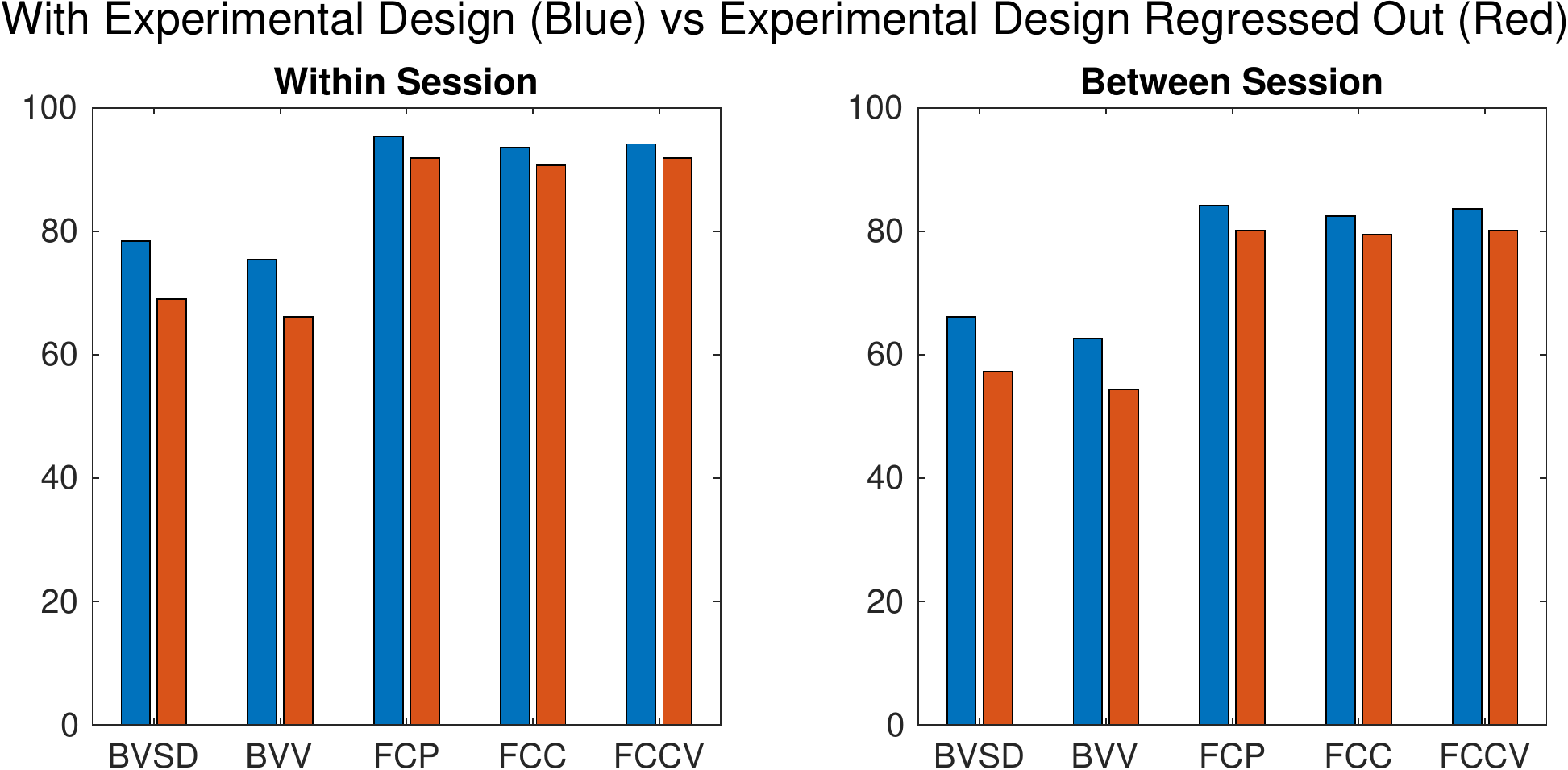}
\caption{ Task classification accuracies for the dataset with experimental design information versus for the dataset with experimental design regressed out. Performance is consistently higher for the dataset containing experimental design. }
\label{task_in_vs_task_out}
\end{figure}

The preprocessing change that resulted in the largest differences in task predictive performance was whether or not the experimental design was regressed out of the BOLD signal (see Figure \ref{task_in_vs_task_out}). Performance decreased when the experimental design was regressed out. 
The performance decrease is greater for BOLD variability features ($\approx 8-10\%$ for BVSD and BVV within-session and between-session) than for FC features ($\approx 2-4\%$ for within-session and $\approx$ 8\% for between-session). 

\subsubsection{Robustness of subject identity classification results across preprocessing options}

Subject identity classification performance was better when using ICA vs. No ICA, but these performance gains were small. The maximum pairwise difference (by feature type) was $\approx 4\%$ for both within-session and between-session prediction.

\begin{figure}[h!]
\includegraphics[width=\textwidth]{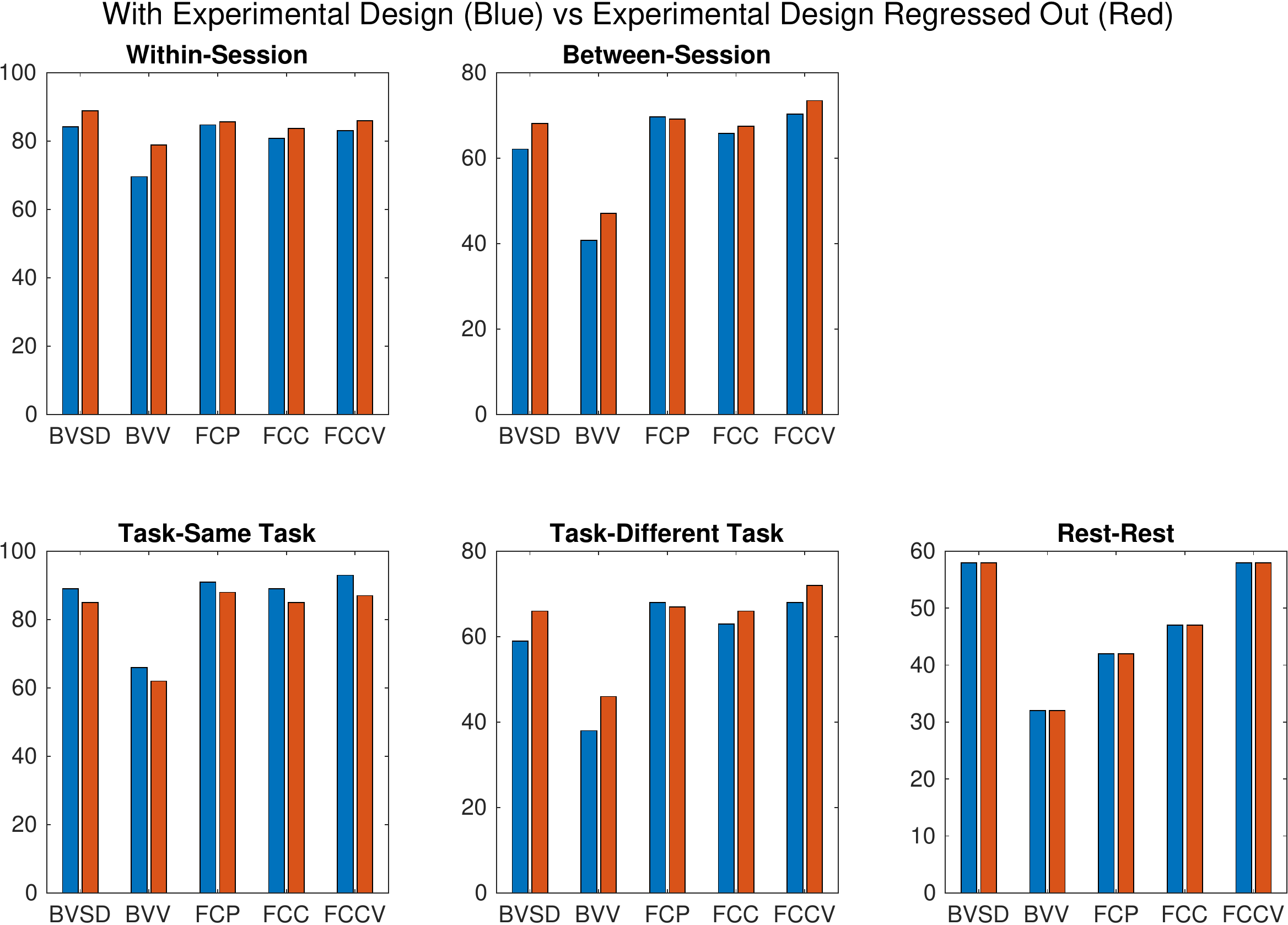}
\caption{ Subject classification accuracies for the dataset with experimental design information versus for the dataset with experimental design regressed out. Average within-session and between-session accuracies are in the first row. The second row decomposes the between-session prediction tasks by training task and test task. }
\label{subj_class_task_in_vs_task_out}
\end{figure}

Including experimental design reduced average predictive performance compared to regressing the experimental design out (see Figure \ref{subj_class_task_in_vs_task_out}). This option affected BV more than FC. BVSD accuracy was reduced by 5\% and 6\% for within-session and between-session prediction, respectively. BVV accuracy was reduced by 9\% and 6\% for within-session and between-session prediction, respectively. FC accuracy was reduced by between 1-3\% for both within-session and between-session prediction. Predictably, the task-same task performance was higher for most features (except for FCCV) when including experimental design. The task-different task performance was worse, and rest-rest performance was unaffected (as there is no experimental design to include). 

\begin{figure}[h!]
\includegraphics[width=\textwidth]{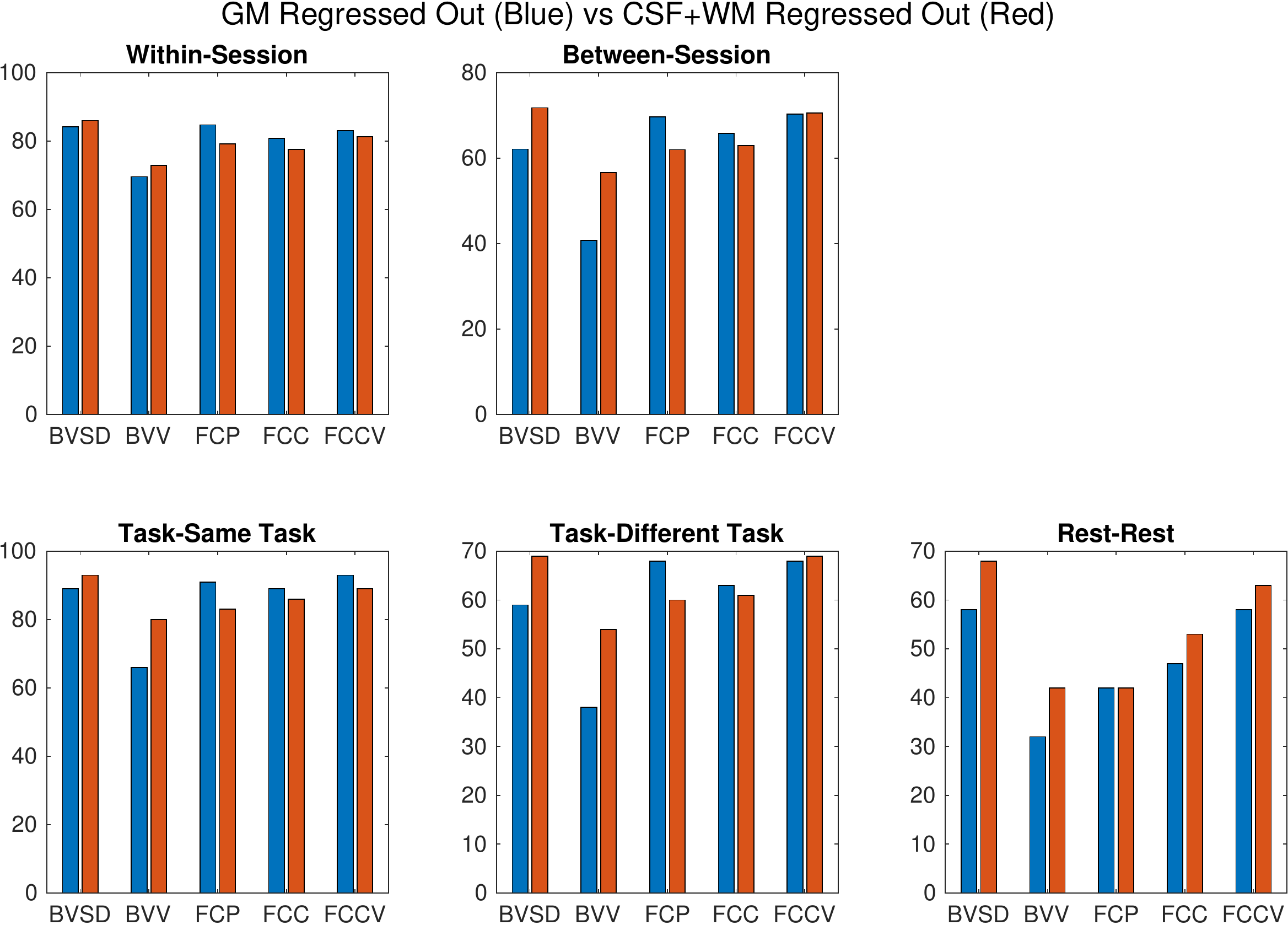}
\caption{ Subject classification accuracies for the dataset with CSF and WM regressed out versus the dataset with MGT regressed out. Average within-session and between-session accuracies are in the first row. The second row decomposes the between-session prediction tasks by training task and test task. }
\label{subj_class_csf_wm_vs_mgt}
\end{figure}

The option to regress out CSF and WM versus MGT also affected subject classification accuracy (see Figure \ref{subj_class_csf_wm_vs_mgt}). For the BV features, MGT regression leads to lower performance within-session, between-session, and for all train/test task pairings that subdivide between-session prediction. For the FC features, the differences in accuracy are reversed to favor MGT regression (except for during rest/rest between-session prediction). 

\subsubsection{Preprocessing Conclusions} 

Exploration of several preprocessing options showed that high predictive performance is largely robust to choice of preprocessing pipeline. Two notable exceptions were that {\it i}) regressing out the experimental design decreased overall task prediction performance and improved overall subject classification performance and {\it ii}) regressing out CSF and WM lead to increased subject classification performance compared to regressing out MGT for BV, but not FC. The performance difference related to {\it ii} suggests that perhaps FC is more sensitive to global motion, whereas BV is more sensitive to physiological sources of noise that aren't motion-related. 

\subsection{ Credible Intervals on Classification Accuracy }

For each model and prediction setting, we report a 95\% credible interval on the classification accuracy. We modeled the classification outcome of the $i$th test instance as a Bernoulli random variable $x_i$ where the probability of a correct classification equals $\theta$ ($p(x_i = 1) = \theta$). Hence, the sum of classification outcomes $X = \sum_{i=1}^N{x_i}$ can be modeled as a Binomial random variable. We used priors to represent the prior belief that the probability of correct classification $\theta$ is near chance. Specifically, we set $\theta \sim \textrm{Beta}(1.25,3)$ (i.e., the mode $\approx 1/9$) for task classification, and $\theta \sim \textrm{Beta}(1.0116,3)$ (i.e., the mode $\approx 1/174$) for subject classification. The posterior distribution of $\theta$ can be computed analytically as $p(\theta| \hat{\alpha}, \hat{\beta}) = \textrm{Beta}(\theta | \hat{\alpha}, \hat{\beta})$ where $\hat{\alpha} = 1 + X$ and $\hat{\beta} = 1 + N - X$. We report the 95\% credible interval on the probability of correct classification as the 2.5\% and 97.5\% percentiles of the posterior distribution over $\theta$. 

\subsection{ Task Descriptions } 
\begin{table*}[h!]
\caption{ Tasks and descriptions. Underlined tasks are those for which we can compute behavioral performance. }
\label{Tasks}
\scriptsize
\hskip-.5cm
\begin{tabularx}{1.1\textwidth}{l X}
{\bf Task} & {\bf Description} \\ \hline \\[-2ex]
\underline{Emotional Pictures} & Subjects see photographs of the screen, one at a time. These photographs appear to the left or right of the center of the screen. The task is to indicate whether the picture is shifted to the left or right relative to green dot in the center of the screen. \\  \\[-2ex]
\underline{Emotional Faces}  & Subjects are presented with male and female faces, one at a time. The task is to determine whether the faces are male or female. There are task conditions for neutral, happy, sad, and fearful faces. \\  \\[-2ex]
 Episodic Memory Encoding &  Subjects see name and face pairings on a screen. The task is to decide whether the name goes well with the face on a 1-4 (poor to well) scale. There are 4 face conditions: young and old faces that are novel or have been repeated during the experiment. \\  \\[-2ex]
\underline{Episodic Memory Retrieval}  & Subjects are asked to remember which names were paired with which faces from the episodic memory encoding task. The task is to indicate whether the face name pairs are the same from the previous task, completely novel, or if the face is repeated, but was not paired with the given name. \\  \\[-2ex]
 \underline{ Go/No-go}  & Subjects are presented with images of single letters. The task is to press a button when the letter is in the set $\{A,B,C,D,E\}$ and not to press the button when the letter is in the set $\{X,Y,Z\}$. \\  \\[-2ex]
  Monetary Incentive Delay  & Subjects are asked to press a button as quickly as possible when a white square (cue) appears on the screen. Participants either win or lose money based on when and how fast they push the button.  \\  \\[-2ex]
 \underline{ Working Memory}  & Subjects are presented with a sequence of letters and switch between the control task and the 2-back memory task. In the control task, subjects are asked to indicate whether the current letter is underlined. In the memory task, subjects are asked to indicate whether the current letter is the same as the one that was presented two letters ago. \\  \\[-2ex]
\underline{ Theory of Mind}  & Subjects are presented with stories and true or false statements about the stories. The task is to indicate whether the statement was true or false. \\  \\[-2ex]
  Resting State  & Subjects are asked to close eyes, relax but stay awake. \\  \\[-2ex]
\end{tabularx}
\end{table*}

\newpage

\section{References}

\bibliographystyle{elsarticle-harv}
\bibliography{fMRIbibfile.bib} 

\end{document}